\documentclass[conference]{IEEEtran}
\IEEEoverridecommandlockouts
\usepackage{cite}
\usepackage{amsmath,amssymb,amsfonts}
\usepackage{amsmath}
\usepackage{algorithmic}
\usepackage{graphicx}
\usepackage{textcomp}
\usepackage{xcolor}
\usepackage{balance} 
\usepackage{array}
\usepackage{float} 
\usepackage{scalerel}
\usepackage{multicol}
\usepackage{esdiff}
\usepackage[hidelinks]{hyperref}
\usepackage[capitalise]{cleveref}
\usepackage{cases}
\usepackage{lipsum}
\usepackage{balance}
\usepackage{mathtools}
\usepackage{comment}
\usepackage{algorithm}
\usepackage{algorithmic}
\usepackage{subcaption}
\usepackage{caption}
\usepackage{graphicx}
\usepackage{placeins}
\usepackage{textcase}
\usepackage{tikz}
\usetikzlibrary{arrows.meta,positioning}

\usepackage[none]{hyphenat} 
\usepackage{microtype}      
\renewcommand{\baselinestretch}{1.03}

\graphicspath{{fig/}}

\def\BibTeX{{\rm B\kern-.05em{\sc i\kern-.025em b}\kern-.08em
		T\kern-.1667em\lower.7ex\hbox{E}\kern-.125emX}}
\begin{document}
	
	\title{A Distributed Quantum Approximate Optimization Algorithm For Unit Commitment\\
		\thanks{This work was supported by the National Science Foundation under Grant ECCS-1944752 and Grant ECCS-2312086.\\}
	}
    
	\author{\IEEEauthorblockN{ Ali Rajabi}
		\IEEEauthorblockA{\textit{Electrical and Computer Engineering}\\
			\textit{Louisiana State University}\\
			Baton Rouge, USA \\
			arajab2@lsu.edu}
		\and
		\IEEEauthorblockN{ Milad Hasanzadeh}
		\IEEEauthorblockA{\textit{Electrical and Computer Engineering}\\
			\textit{Louisiana State University}\\
			Baton Rouge, USA \\
			mhasa42@lsu.edu}
		\and
		\IEEEauthorblockN{ Amin Kargarian}
		\IEEEauthorblockA{\textit{Electrical and Computer Engineering}\\
			\textit{Louisiana State University}\\
			Baton Rouge, USA \\
			kargarian@lsu.edu}
	}
	
	\maketitle

\begin{abstract}
This paper presents a distributed quantum approximate optimization algorithm (DQAOA)-enabled three-block alternating direction method of multipliers (ADMM) framework for unit commitment (UC). The relaxed commitment and dispatch variables are solved in a continuous quadratic programming block, while the binary commitment block is formulated as a quadratic unconstrained binary optimization (QUBO) problem. The DQAOA interface allows this QUBO to be solved using brute-force enumeration, monolithic QAOA, or distributed QAOA, while the remaining ADMM updates are kept unchanged. In the distributed mode, the logical commitment qubits are allocated across multiple capacity-constrained quantum processing units (QPU), avoiding the requirement that the complete binary problem fits on a single device. The framework is evaluated on a five-unit UC instance containing 15 binary variables. All three solver modes reduce the ADMM primal residual below a certain tolerance and recover the same commitment schedule, dispatch, and operating cost. The results demonstrate solution consistency across the three solver modes and the multi-QPU capacity accommodation provided by the distributed QAOA method.
\end{abstract}


\section{Introduction}
\label{sec:introduction}

Unit commitment (UC) is a fundamental scheduling problem in power system operation and electricity markets. It determines which generating units should be committed and how their outputs should be scheduled to meet system demand while satisfying operational requirements at minimum cost~\cite{padhy2004unit}. Because UC combines binary on/off decision variables with continuous dispatch variables, it is commonly formulated as a mixed-integer optimization problem~\cite{carrion2006computationally}. The computational complexity of UC increases as the number of generating units, scheduling periods, and operating constraints grows. Consequently, the development of efficient UC solution methods remains an active research topic~\cite{hasanzadeh2026survey}.

UC has traditionally been addressed using classical optimization
techniques. Its binary decision component is commonly handled through methods such as branch-and-bound, branch-and-cut, and
decomposition algorithms, as well as heuristic and metaheuristic search procedures~\cite{hasanzadeh2026survey}. Although these approaches have been successfully applied to many UC formulations, their computational burden can increase considerably as the problem size and operational complexity grow. Therefore, decomposition methods have become particularly useful because they separate binary commitment decisions from continuous dispatch calculations and coordinate them via iterative updates. This structure also creates an opportunity to replace the binary optimization procedure while retaining efficient classical solvers for the continuous components. Among the available decomposition frameworks, three-block alternating direction method of multipliers (three-block ADMM) is particularly suitable for this purpose because it coordinates continuous and binary variable blocks within a unified iterative procedure~\cite{gambella2020multiblock}.

In a three-block UC formulation, the relaxed commitment and dispatch variables are handled in the first block through a quadratic program (QP) with continuous variables. The binary commitment decisions are isolated in the second block, while the third block updates a slack variable to enforce consensus between the relaxed and binary commitment representations. This structure allows the solution method used for the binary block to be changed without modifying the remaining UC and ADMM calculations. A hybrid quantum--classical UC method based on this decomposition was previously presented in~\cite{mahroo2022hybrid}.

The second block can be expressed as a quadratic unconstrained binary optimization (QUBO) problem, which provides a standard representation for many combinatorial optimization models \cite{glover2018tutorial}. A QUBO can be transformed into an Ising cost Hamiltonian that is compatible with gate-based quantum optimization algorithms \cite{lucas2014ising}. Among these algorithms, the quantum approximate optimization algorithm (QAOA) is designed specifically for discrete optimization and constructs a parameterized circuit by alternating problem-dependent cost operations with mixer operations \cite{farhi2014quantum}. A classical optimizer adjusts the circuit parameters to reduce the expected cost and increase the likelihood of observing low-cost binary solutions. QAOA and its variants have consequently received significant attention for near-term combinatorial optimization~\cite{zhou2020quantum}.

Quantum approaches to UC have included annealing-based, variational circuit-based, and hybrid decomposition methods, as reviewed in~\cite{hasanzadeh2026survey}. The hybrid method in
\cite{mahroo2022hybrid} demonstrated that a quantum routine can be used for the binary UC block while the continuous calculations remain classical. However, a monolithic quantum implementation requires the complete circuit and all logical data qubits to be accommodated by a single quantum processing unit (QPU). This requirement can become restrictive because near-term devices have limited qubit capacity, connectivity, and circuit fidelity \cite{preskill2018quantum}.

Distributed quantum computing provides an alternative in which a quantum circuit is implemented across multiple connected QPUs~\cite{cuomo2020towards}. Circuit operations involving qubits located on different QPUs can be implemented through distributed gate protocols and classical communication \cite{diadamo2021distributed}. Distributed QAOA extends this idea to combinatorial optimization by assigning logical variables and circuit operations across multiple quantum devices \cite{yue2023local}. Its main practical motivation is to use the
combined qubit capacities of multiple QPUs rather than requiring the complete problem to fit on a single device.

A recent study integrated a distributed variational quantum eigensolver (DVQE) into the binary block of a three-block ADMM formulation for UC~\cite{hasanzadeh2026distributed}. Furthermore, the DQAOA framework introduced in~\cite{rajabi2026distributed} provides a common interface for solving QUBO problems using brute-force enumeration, monolithic QAOA, or distributed QAOA. The framework assigns logical data qubits based on the available QPU capacities and supports the distributed realization of interactions that span QPU boundaries. However, its integration as the binary solution routine within a complete three-block ADMM workflow for UC has not yet been evaluated in detail.

This paper develops a DQAOA-enabled three-block ADMM framework for UC. At each ADMM iteration, the first block updates the relaxed commitment and dispatch variables. These updated values, together with the current slack and dual multiplier variables, are then used to construct the QUBO objective associated with the binary commitment block. The resulting QUBO is passed to the common DQAOA interface and solved using brute-force enumeration, monolithic QAOA, or distributed QAOA. The selected binary commitment solution is returned to the ADMM loop, after which the slack variable and dual multiplier updates are performed in the same manner for all three solver modes. Accordingly, the UC formulation, ADMM parameters, QUBO construction procedure, and convergence criterion remain unchanged across the three modes, while only the solution method for the binary block varies.

The main contributions of this work are summarized as follows:
\begin{itemize}
    \item The DQAOA package is integrated as a selectable solver for the iteration-dependent binary block within a three-block ADMM formulation for UC.

    \item Brute force, monolithic QAOA, and distributed QAOA are evaluated using identical UC data, QUBO objective coefficients, ADMM settings, and convergence criteria.

    \item The distributed implementation allocates the logical commitment qubits across multiple capacity-constrained QPUs while recovering the same commitment schedule, dispatch, and operating cost as the monolithic and brute-force modes for the tested case.
\end{itemize}

\section{Problem Formulation}
\label{sec:problem_formulation}

This section introduces the unit commitment model considered in this work. It describes its decomposition into three coordinated subproblems: (i) a convex quadratic (QP) problem for relaxed binary commitment variables and continuous generation dispatch, (ii) a  QUBO problem to determine the status of binary variables, and (iii) a linear optimization problem based on slack variable update with closed-form solution to enforce UC feasibility and satisfy consensus constraint. The adopted three-block ADMM formulation follows the general construction
in~\cite{hasanzadeh2026distributed}, whereas the binary subproblem in the present work is solved using different modes including brute force, monolithic QAOA, or distributed QAOA. The remaining ADMM blocks and their associated constraints are kept unchanged across all solver modes.

\subsection{Unit Commitment Model}
\label{subsec:uc_model}

Consider a scheduling horizon containing $T$ time periods and a system
with $N$ generating units. The binary variable $y_{i,t}$ indicates whether
unit $i$ is committed during period $t$, and $p_{i,t}$ denotes its power
output. The simplified UC problem is formulated as follows:
\begin{subequations}\label{eq:uc}
\begin{align}
\min_{y_{i,t},\,p_{i,t}}\quad
&
\sum_{t=1}^{T}\sum_{i=1}^{N}
\left(
A_i y_{i,t}
+
B_i p_{i,t}
+
C_i p_{i,t}^{2}
\right),
\label{eq:uc_objective}
\\
\text{s.t.}\quad
&
\sum_{i=1}^{N}p_{i,t}=L_t,
\qquad \forall t,
\label{eq:uc_balance}
\\
&
p_i^{\min}y_{i,t}
\leq p_{i,t}
\leq
p_i^{\max}y_{i,t},
\qquad \forall i,t,
\label{eq:uc_capacity}
\\
&
y_{i,t}\in\{0,1\},
\qquad \forall i,t.
\label{eq:uc_binary}
\end{align}
\end{subequations}
Here, $L_t$ denotes the demand during period $t$.
The coefficients $A_i$, $B_i$, and $C_i$ represent the fixed, linear, and quadratic operating cost parameters of unit $i$, respectively. The quantities $p_i^{\min}$ and $p_i^{\max}$ specify the lower and upper generation limits of generating unit $i$. 
The objective function in \eqref{eq:uc_objective} minimizes the total operating cost over the scheduling horizon. Constraint~\eqref{eq:uc_balance} maintains the supply--demand balance, while constraint~\eqref{eq:uc_capacity} couples the continuous dispatch variable to the
binary commitment variable. When $y_{i,t}=0$, the corresponding generation output is
forced to zero, while $y_{i,t}=1$ indicates that the unit is allowed to operate inside its
admissible output range.

\subsection{Three-Block ADMM Decomposition}
\label{subsec:three_block_admm}

To separate the continuous generation variables from the binary
commitment decisions, the binary variable $y_{i,t}$ in
\eqref{eq:uc_binary} is first relaxed over the interval $[0,1]$.
An auxiliary binary commitment variable $z_{i,t}$ is then introduced to preserve the binary requirement. The relaxed and binary representations are coordinated through a slack variable $r_{i,t}$ and the associated consensus equality constraint. Following the three-block construction in~\cite{hasanzadeh2026distributed}, the binary constraint in
\eqref{eq:uc_binary} is replaced by
\begin{subequations}\label{eq:consensus}
\begin{align}
0 \leq y_{i,t}\leq 1,
&\qquad \forall i,t,
\label{eq:y_relaxed}
\\
z_{i,t}\in\{0,1\},
&\qquad \forall i,t,
\label{eq:z_binary}
\\
r_{i,t}\in\mathbb{R},
&\qquad \forall i,t,
\label{eq:r_domain}
\\
y_{i,t}-z_{i,t}+r_{i,t}=0,
&\qquad \forall i,t.
\label{eq:consensus_constraint}
\end{align}
\end{subequations}
In this decomposition, $y_{i,t}$ remains as the continuous relaxed commitment, whereas $z_{i,t}$ carries the binary commitment requirement. The variable $r_{i,t}$ supports coordination between these two representations through \eqref{eq:consensus_constraint}.

Let $\lambda_{i,t}$ denote the Lagrange multiplier associated with the
consensus constraint, $\rho>0$ denote the augmented Lagrangian penalty
coefficient, and $\beta>0$ denote the regularization parameter applied to
$r$. Defining the continuous variables as
$\Delta=\{y,p\}$, the augmented Lagrangian is expressed as
\begin{align}
\mathcal{L}_{\rho}(\Delta,z,r,\lambda)
={}&
\sum_{t=1}^{T}\sum_{i=1}^{N}
\left(
A_i y_{i,t}
+
B_i p_{i,t}
+
C_i p_{i,t}^{2}
\right)
\nonumber\\
&+
\sum_{t=1}^{T}\sum_{i=1}^{N}
\lambda_{i,t}
\left(
y_{i,t}-z_{i,t}+r_{i,t}
\right)
\nonumber\\
&+
\frac{\rho}{2}
\left\|
y_{i,t}-z_{i,t}+r_{i,t}
\right\|_{2}^{2}
+
\frac{\beta}{2}
\left\|
r_{i,t}
\right\|_{2}^{2}
\label{eq:augLag}
\end{align}
The optimization remains subject to the power balance and generation limit
constraints in \eqref{eq:uc_balance}--\eqref{eq:uc_capacity}, together
with the domains specified in \eqref{eq:consensus}. The three-block ADMM
procedure updates the continuous variables, binary variables, and slack
variables sequentially, followed by a correction of the Lagrange
multipliers.

\paragraph{Block 1: Continuous variables update}
With $z$, $r$, and $\lambda$ fixed at their values from the preceding
iteration, the relaxed commitment and dispatch variables are updated
according to
\begin{equation}
\Delta^{(k)}
=
\arg\min_{\Delta}
\mathcal{L}_{\rho}
\left(
\Delta,
z^{(k-1)},
r^{(k-1)},
\lambda^{(k-1)}
\right),
\label{eq:block1_cont}
\end{equation}
subject to \eqref{eq:uc_balance}--\eqref{eq:uc_capacity} and
$0\leq y_{i,t}\leq1$. This block is a convex QP in the relaxed commitment
and dispatch variables.

\paragraph{Block 2: Binary commitment update}
After obtaining $\Delta^{(k)}$, the binary commitment vector is determined
from
\begin{equation}
z^{(k)}
=
\arg\min_{z\in\{0,1\}^{NT}}
\mathcal{L}_{\rho}
\left(
\Delta^{(k)},
z,
r^{(k-1)},
\lambda^{(k-1)}
\right).
\label{eq:block2_cont}
\end{equation}
The terms of \eqref{eq:block2_cont} that depend on $z$ can be collected in
the standard QUBO form
\begin{equation}
\min_{z\in\{0,1\}^{NT}}
\quad
z^{\mathsf T}Q^{(k)}z
+
\left(q^{(k)}\right)^{\mathsf T}z
+
c^{(k)},
\label{eq:block2_qubo}
\end{equation}
where $Q^{(k)}$, $q^{(k)}$, and $c^{(k)}$ are obtained from the current
ADMM variables and parameters. Equation~\eqref{eq:block2_qubo} is the binary subproblem supplied to the selected solver. In this work, the QUBO problem is solved using brute-force enumeration, monolithic QAOA, or distributed QAOA. Therefore, only the solution method used in Block 2 changes among the three modes.

\paragraph{Block 3: Slack variable update}
Once the continuous and binary variables have been updated, the third block
computes
\begin{equation}
r^{(k)}
=
\arg\min_{r}
\mathcal{L}_{\rho}
\left(
\Delta^{(k)},
z^{(k)},
r,
\lambda^{(k-1)}
\right).
\label{eq:block3_cont}
\end{equation}
Since this subproblem is unconstrained and quadratic in $r$, its solution
is obtained analytically as
\begin{equation}
r_{i,t}^{(k)}
=
-
\frac{
\lambda_{i,t}^{(k-1)}
+
\rho
\left(
y_{i,t}^{(k)}
-
z_{i,t}^{(k)}
\right)
}{
\beta+\rho
}.
\label{eq:slack_update}
\end{equation}

\paragraph{Dual variable update}
The Lagrange multipliers are subsequently adjusted using the consensus
violation remaining after the three primal updates:
\begin{equation}
\lambda_{i,t}^{(k)}
=
\lambda_{i,t}^{(k-1)}
+
\frac{\rho}{2}
\left(
y_{i,t}^{(k)}
-
z_{i,t}^{(k)}
+
r_{i,t}^{(k)}
\right).
\label{eq:dual_update}
\end{equation}

The ADMM iterations continue until the primal consensus residual
\begin{equation}
\varepsilon_{\mathrm{pri}}^{(k)}
=
\left\|
y^{(k)}
-
z^{(k)}
+
r^{(k)}
\right\|_{2}
\label{eq:primal_residual}
\end{equation}
falls below a certain tolerance or the maximum number of ADMM iterations is reached.

\section{Hybrid DQAOA Framework}
\label{sec:dqaoa_framework}

At each ADMM iteration, the binary commitment block generates a QUBO objective whose coefficients depend on the current relaxed commitment, slack, and dual multiplier variables. The developed DQAOA package solves this QUBO problem using brute force, monolithic QAOA, or distributed QAOA. All three modes receive the
same QUBO coefficients, while the continuous, slack, and dual updates
remain unchanged.

\subsection{QUBO Encoding and Monolithic QAOA}
\label{subsec:monolithic_qaoa}

At iteration $k$, Block~2 is expressed as
\begin{equation}
\begin{gathered}
F^{(k)}(\boldsymbol{z})
=
\boldsymbol{z}^{\mathsf T}\boldsymbol{H}^{(k)}\boldsymbol{z}
+
\left(\boldsymbol{f}^{(k)}\right)^{\mathsf T}\boldsymbol{z}
+
c_0^{(k)},
\\
\boldsymbol{z}\in\{0,1\}^{n}.
\end{gathered}
\label{eq:iteration_qubo}
\end{equation}
where $n=NT$ is the number of binary commitment variables. Using the standard binary-to-Ising mapping $\widehat{z}_j=(I-Z_j)/2$, the QUBO objective is transformed into a cost Hamiltonian expressed in terms of Pauli-$Z$ operators.
\begin{equation}
H_C^{(k)}
=
\alpha^{(k)}I
+
\sum_{j=1}^{n}h_j^{(k)}Z_j
+
\sum_{j<m}J_{jm}^{(k)}Z_jZ_m,
\label{eq:cost_hamiltonian}
\end{equation}
where $I$ is the identity operator, $Z_j$ is the Pauli-$Z$ operator acting on qubit $j$, and $\alpha^{(k)}$, $h_j^{(k)}$, and $J_{jm}^{(k)}$ are obtained from the QUBO coefficients.

QAOA alternates the unitaries generated by the cost and the mixer Hamiltonians. The mixer Hamiltonian is
\begin{equation}
H_M=\sum_{j=1}^{n}X_j,
\label{eq:mixer_hamiltonian}
\end{equation}
where $X_j$ is the Pauli-$X$ operator acting on qubit $j$. At depth $p$, the variational state is
\begin{equation}
|\psi_p^{(k)}(\boldsymbol{\gamma},\boldsymbol{\beta})\rangle
=
\prod_{\ell=1}^{p}
e^{-\mathrm{i}\beta_\ell H_M}
e^{-\mathrm{i}\gamma_\ell H_C^{(k)}}
|+\rangle^{\otimes n}.
\label{eq:qaoa_state}
\end{equation}
A classical optimizer updates the variational parameters of the quantum circuit ($\boldsymbol{\gamma}$ and
$\boldsymbol{\beta}$) to minimize the expectation value of the cost
Hamiltonian
\begin{equation}
\begin{aligned}
E^{(k)}(\boldsymbol{\gamma},\boldsymbol{\beta})
={}&
\langle
\psi_p^{(k)}(\boldsymbol{\gamma},\boldsymbol{\beta})
|
H_C^{(k)}
|
\psi_p^{(k)}(\boldsymbol{\gamma},\boldsymbol{\beta})
\rangle.
\end{aligned}
\label{eq:qaoa_objective}
\end{equation}
The optimized circuit is then sampled to obtain candidate commitment
bitstrings. In monolithic QAOA, all $n$ logical data qubits are placed on
one QPU; therefore, the capacity of that QPU must satisfy
$c_{\mathrm{mono}}\geq n$.

\subsection{Capacity-Constrained Distributed QAOA}
\label{subsec:dqaoa}

DQAOA preserves the complete QUBO objective and the circuit depth of QAOA structure introduced in Section~\ref{subsec:monolithic_qaoa}, but
distributes the logical data qubits across multiple QPUs. Thus, Block~2 is not separated into independent QUBO subproblems; instead, the circuit representing the original cost Hamiltonian is implemented over several capacity-constrained QPUs.

Let $\mathcal{V}=\{1,\ldots,n\}$ denote the binary variable indices and let $c_m$ be the available data qubit capacity of QPU $m$. The assigned variable subsets $\mathcal{V}_m$ satisfy
\begin{equation}
\begin{aligned}
&\bigcup_{m=1}^{M}\mathcal{V}_m=\mathcal{V},
\qquad
\mathcal{V}_m\cap\mathcal{V}_{m'}=\emptyset,
\quad m\neq m',
\\
&|\mathcal{V}_m|\leq c_m,\quad \forall m,
\qquad
\sum_{m=1}^{M}c_m\geq n.
\end{aligned}
\label{eq:qpu_allocation}
\end{equation}
Each binary variable is mapped to one logical data qubit. Consequently, monolithic QAOA requires a single QPU with at least $n$ data qubits, whereas DQAOA accommodates the same number of data qubits using the combined capacities of several QPUs.

\begin{figure}[t]
\centering

\subfloat[Monolithic QAOA.]{
    \includegraphics[width=0.99\columnwidth]
    {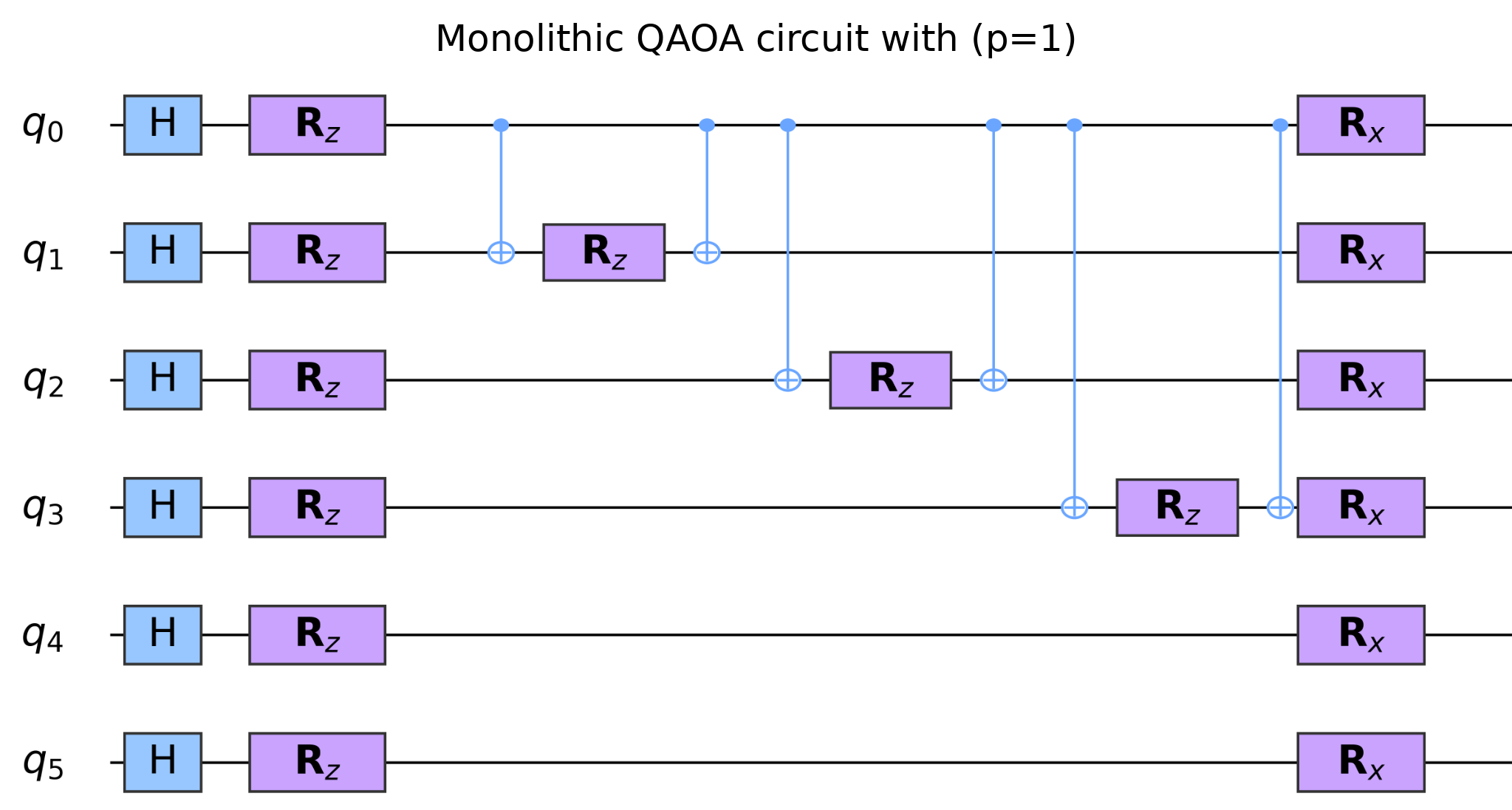}
    \label{fig:monolithic_qaoa_circuit}
}

\vspace{1mm}

\subfloat[Distributed QAOA.]{
    \includegraphics[width=0.99\columnwidth]
    {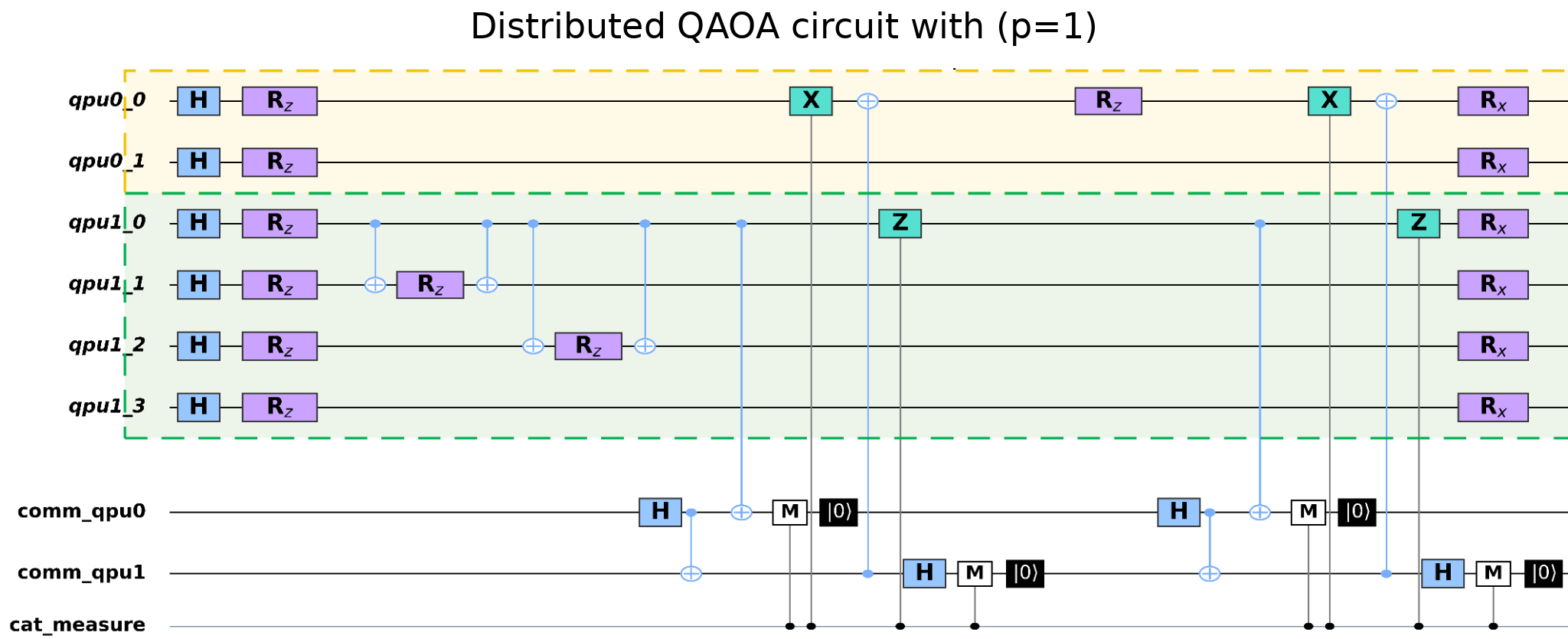}
    \label{fig:distributed_qaoa_circuit}
}

\caption{Monolithic and distributed QAOA circuits
for a six-variable QUBO.}
\label{fig:qaoa_circuit_comparison}
\end{figure}

After allocation, each quadratic cost term
$J_{jm}^{(k)}Z_jZ_m$ is classified according to the locations of its two qubits. If both qubits are assigned to the same QPU, the interaction is implemented locally. If they are assigned to different QPUs, the corresponding operation is realized through the TeleGate-based mechanism of the DQAOA package~\cite{rajabi2026distributed}. TeleGate uses communication qubits, intermediate measurements, and classically conditioned corrections to reproduce the required cross-QPU interaction while preserving the logical data qubits. The mixer operations remain local because they act independently on each data qubit. All QPUs use the same global QAOA parameters $\boldsymbol{\gamma}$ and $\boldsymbol{\beta}$, and the classical optimizer minimizes the same expected QUBO cost as in monolithic QAOA. The main benefit is therefore multi-QPU qubit capacity accommodation, rather than runtime acceleration, because cross-QPU operations introduce additional communication and circuit overhead. Fig.~\ref{fig:qaoa_circuit_comparison} compares monolithic and distributed QAOA circuits for a six-variable QUBO. The monolithic circuit uses a single QPU, whereas the distributed circuit distributes the logical data qubits across two QPUs and uses communication qubits for cross-QPU interactions.

\subsection{Integration Within the Three-Block ADMM}
\label{subsec:dqaoa_admm_integration}

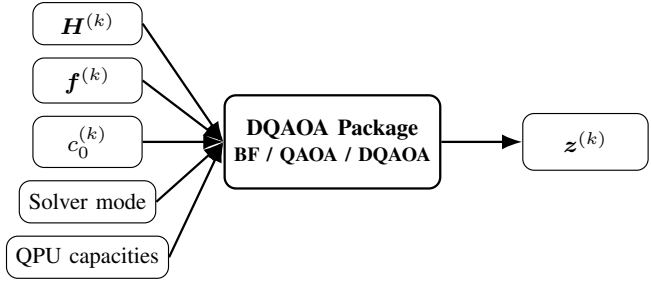
\begin{figure}[t]
\centering
\resizebox{0.96\columnwidth}{!}{%
\begin{tikzpicture}[
    input/.style={
        rectangle,
        rounded corners,
        draw,
        align=center,
        minimum width=1.35cm,
        minimum height=0.52cm,
        font=\footnotesize
    },
    solver/.style={
        rectangle,
        rounded corners,
        draw,
        thick,
        align=center,
        minimum width=2.6cm,
        minimum height=1.15cm,
        font=\footnotesize\bfseries
    },
    output/.style={
        rectangle,
        rounded corners,
        draw,
        align=center,
        minimum width=1.55cm,
        minimum height=0.7cm,
        font=\footnotesize
    },
    arrow/.style={-{Latex[length=2.5mm]}, thick},
    node distance=1.5mm and 7mm
]

\node[input] (H) {$\boldsymbol{H}^{(k)}$};
\node[input, below=of H] (f) {$\boldsymbol{f}^{(k)}$};
\node[input, below=of f] (c) {$c_0^{(k)}$};
\node[input, below=of c] (mode) {Solver mode};
\node[input, below=of mode] (cap) {QPU capacities};

\node[solver, right=10mm of c] (solver)
{DQAOA Package\\
\scriptsize BF / QAOA / DQAOA};

\node[output, right=10mm of solver] (z)
{$\boldsymbol{z}^{(k)}$};

\draw[arrow] (H.east) -- (solver.west);
\draw[arrow] (f.east) -- (solver.west);
\draw[arrow] (c.east) -- (solver.west);
\draw[arrow] (mode.east) -- (solver.west);
\draw[arrow] (cap.east) -- (solver.west);
\draw[arrow] (solver.east) -- (z.west);

\end{tikzpicture}%
}
\caption{Integration of the DQAOA package as the selectable Block~2
binary solver within the three-block ADMM framework.}
\label{fig:dqaoa_solver_interface}
\end{figure}

The developed DQAOA package is integrated into Block~2 as the binary solver of the three-block ADMM framework. At iteration $k$,
Block~1 first updates the continuous relaxed commitment and dispatch variables. These values, together with the current slack and dual multiplier variables, determine the QUBO coefficients $\boldsymbol{H}^{(k)}$, $\boldsymbol{f}^{(k)}$, and $c_0^{(k)}$ in \eqref{eq:iteration_qubo}. As illustrated in Fig.~\ref{fig:dqaoa_solver_interface}, the coefficients and selected solver mode are passed to the DQAOA interface, which returns the binary commitment vector $\boldsymbol{z}^{(k)}$ to the ADMM loop. For Distributed QAOA mode, QPU capacities are also provided to determine the structure of variable-to-QPU allocation.

In brute-force mode, the binary search space is enumerated to obtain an
exact Block~2 solution for sufficiently small instances. In monolithic QAOA and Distributed QAOA modes, the variational parameters $\boldsymbol{\gamma}$ and $\boldsymbol{\beta}$ are trained by repeatedly executing the corresponding parametrized quantum circuit, estimating the expected QUBO cost and updating the parameters using a classical Adam optimizer. Once the training
stage is complete, the optimized circuit is executed using a certain number of final shots to produce sampled bitstrings and their frequencies. The measured outcomes are decoded into the logical ordering of the commitment variables. In Distributed QAOA mode, measurement bits associated with communication bits and intermediate TeleGate operations are excluded during this decoding step. Let $\mathcal{S}^{(k)}$ denote the set of sampled distinct logical bitstrings obtained from the final circuit sampling.
Each sampled bitstring is evaluated using the current Block~2 QUBO
objective, and the best observed solution is selected as
\begin{equation}
\boldsymbol{z_{opt}}^{(k)}
=
\arg\min_{\boldsymbol{z}\in\mathcal{S}^{(k)}}
F^{(k)}(\boldsymbol{z}).
\label{eq:best_sampled_bitstring}
\end{equation}
The resulting commitment vector $\boldsymbol{z_{opt}}^{(k)}$ is returned to the ADMM loop and used in the subsequent slack variable and dual multiplier updates. Therefore, the developed package replaces only the solution procedure for Block~2, while the UC formulation and the remaining ADMM operations are identical across brute-force, monolithic QAOA, and Distributed QAOA modes.

\begin{algorithm}[t]
\footnotesize
\setlength{\algorithmicindent}{0.99em}
\caption{DQAOA-Enabled Three-Block ADMM}
\label{alg:dqaoa_admm}
\begin{algorithmic}[1]
\STATE \textbf{Initialize:}
$\Delta^{(0)}$, $z^{(0)}$, $r^{(0)}$, $\lambda^{(0)}$,
ADMM parameters, solver mode, and QPU capacities.
\STATE Set ADMM iteration: $k\gets1$.
\WHILE{stopping criterion is not satisfied}
    \STATE \textbf{Block 1:}
    Solve \eqref{eq:block1_cont} for
    $(y^{(k)},p^{(k)})$.
    \STATE \textbf{Block 2:}
    Construct
    $(\boldsymbol{H}^{(k)},\boldsymbol{f}^{(k)},c_0^{(k)})$
    for the QUBO.
    \IF{mode is brute-force}
        \STATE Enumerate the binary space and obtain the
        bitstring $z^{(k)}$ with minimum QUBO cost.
    \ELSE
        \IF{mode is Monolithic QAOA}
            \STATE Allocate the logical data qubits on a single QPU according to \eqref{eq:qpu_allocation} with $M = 1$.
            \STATE Construct the monolithic QAOA circuit.
        \ENDIF
    
        \IF{mode is Distributed QAOA}
            \STATE Allocate the logical data qubits according to
            \eqref{eq:qpu_allocation}.
            \STATE Construct the distributed QAOA circuit.
        \ENDIF
        \STATE Optimize
        $(\boldsymbol{\gamma},\boldsymbol{\beta})$ to minimize $E^{(k)}(\boldsymbol{\gamma},\boldsymbol{\beta})$ using \eqref{eq:qaoa_objective}.
        \STATE Sample the trained circuit and decode the logical
        bitstrings.
        \STATE Select $z_{opt}^{(k)}$ using
        \eqref{eq:best_sampled_bitstring}.
    \ENDIF
    \STATE \textbf{Block 3:}
    Update $r^{(k)}$ using \eqref{eq:slack_update}.
    \STATE \textbf{Dual:}
    Update $\lambda^{(k)}$ using \eqref{eq:dual_update}.
    \STATE Compute the primal residual in
    \eqref{eq:primal_residual}.
    \STATE Set $k\gets k+1$.
\ENDWHILE
\STATE \textbf{Return:}
$(y^{(k)},p^{(k)},z^{(k)},r^{(k)},\lambda^{(k)})$.
\end{algorithmic}
\end{algorithm}

\section{Case Study}
\label{sec:case_study}

\subsection{Test System and Settings}
\label{subsec:test_system}

The proposed framework is evaluated on a five-unit, three-period UC instance with the demand profile
\begin{equation}
\boldsymbol{L}=[60,\;130,\;280]\ {\rm MW}.
\label{eq:demand_profile}
\end{equation}
The resulting problem contains $n=NT=15$ binary variables.
The same UC data, ADMM parameters, initialization, stopping tolerance, and Block~2 QUBO coefficients are used across the brute-force, monolithic QAOA, and distributed QAOA modes. Brute-force provides the exact reference for small problems. Monolithic QAOA represents all 15 commitment variables on one QPU, whereas distributed QAOA distributes the corresponding logical data qubits across multiple
capacity-constrained QPUs. The primal residual tolerance is set to
$10^{-3}$.

\begin{table}[t]
\centering
\caption{Five-Unit System Data}
\label{tab:5unit-data}
\footnotesize
\setlength{\tabcolsep}{3.2pt}
\resizebox{\columnwidth}{!}{%
\begin{tabular}{c c c c c c}
\hline
Unit
& $A_i$ (\$)
& $B_i$ (\$/MW)
& $C_i$ (\$/MW$^2$)
& $P_i^{\min}$ (MW)
& $P_i^{\max}$ (MW)
\\
\hline
1 & 185 & 25.0 & 0.0580 & 10 & 80  \\
2 & 170 & 21.0 & 0.0160 & 5  & 50  \\
3 & 275 & 22.0 & 0.0165 & 5  & 60  \\
4 & 220 & 24.5 & 0.0110 & 10 & 85  \\
5 & 150 & 24.7 & 0.0070 & 20 & 110 \\
\hline
\end{tabular}%
}
\end{table}

\subsection{Solution Consistency and ADMM Convergence}
\label{subsec:case_results}

\begin{figure}[t]
\centering

\subfloat[Monolithic QAOA]{
    \includegraphics[width=0.95\columnwidth]
    {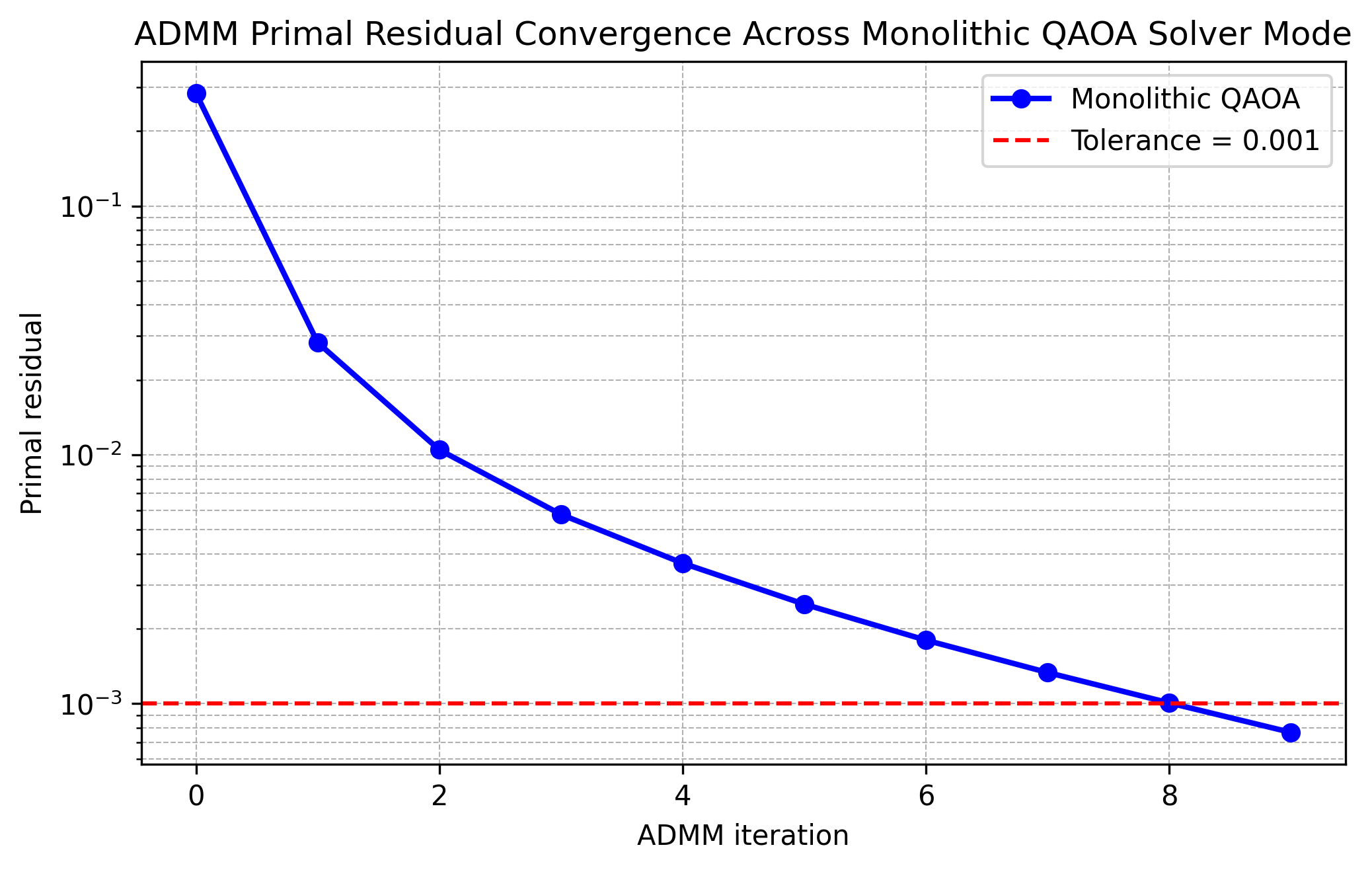}
    \label{fig:residual_mono}
}

\vspace{2mm}

\subfloat[Distributed QAOA]{
    \includegraphics[width=0.95\columnwidth]
    {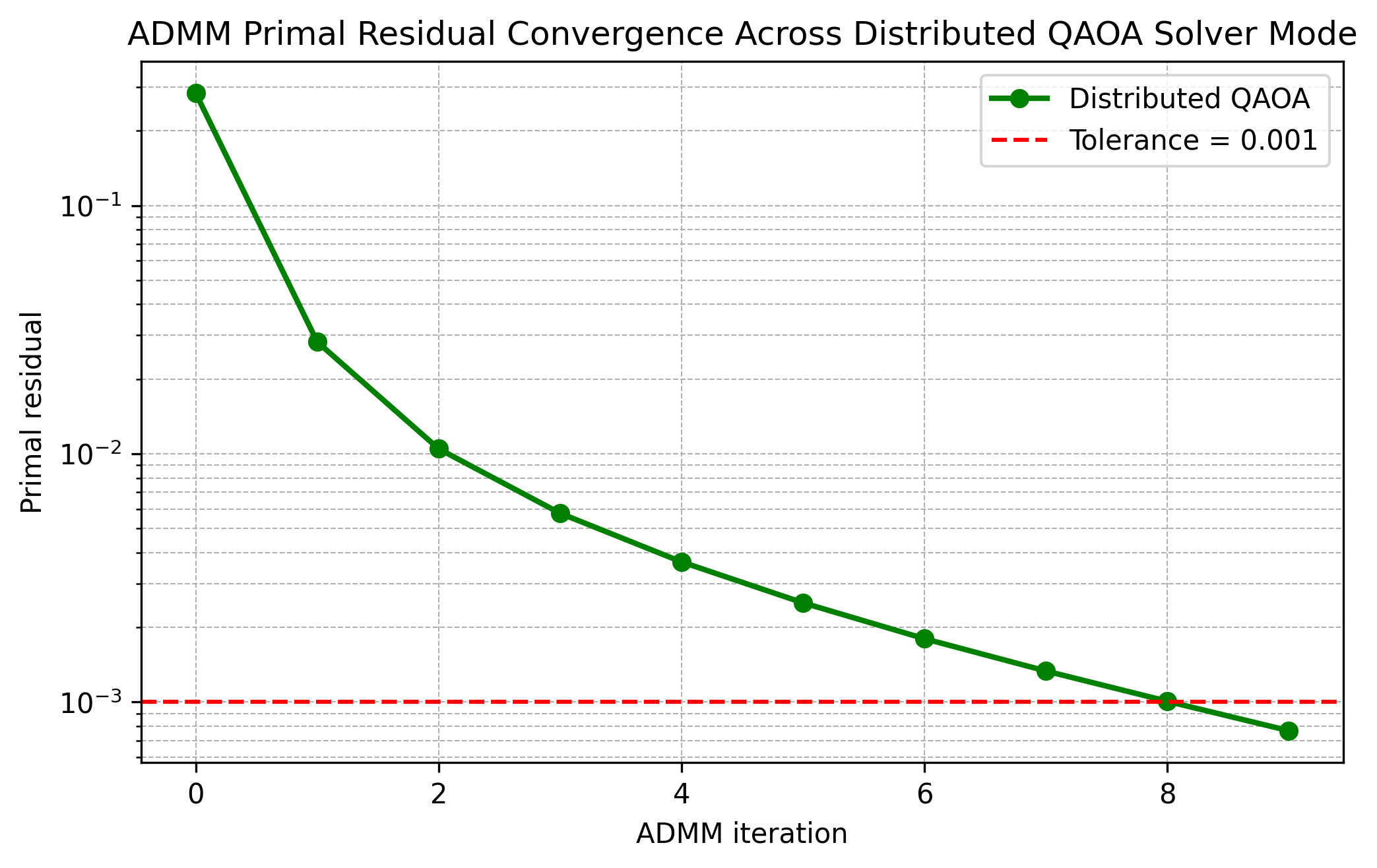}
    \label{fig:residual_dqaoa}
}

\caption{ADMM primal residual convergence for the five-unit,
three-period UC case using (a) monolithic QAOA, and
(b) distributed QAOA as the Block~2 solver. The dashed line denotes the
convergence tolerance of $10^{-3}$.}
\label{fig:residual_comparison}
\end{figure}

Fig.~\ref{fig:residual_comparison} indicates the ADMM primal residual for the solver quantum modes of Block~2, where the primal residual decreases below a tolerance of $10^{-3}$, indicating satisfaction of the consensus constraint. Although the quantum-based modes use finite-shot circuit evaluations, both monolithic QAOA and Distributed QAOA modes converge to the same commitment schedule as the brute-force reference. All three modes recover the commitment matrix
\begin{equation}
\boldsymbol{z}^{\star}
=
\begin{bmatrix}
0&0&1&0&0\\
0&1&0&0&1\\
0&1&1&1&1
\end{bmatrix},
\label{eq:optimal_commitment}
\end{equation}
where each row corresponds to one scheduling period and each column
corresponds to one generating unit. Reading the commitment matrix row by row, with each row corresponding to one scheduling period, gives the bitstring
\begin{equation}
\boldsymbol{z}_{\mathrm{opt}}^{\star}
=
001000100101111,
\label{eq:optimal_bitstring}
\end{equation}
and the associated dispatch is
\begin{equation}
\boldsymbol{p}^{\star}
=
\begin{bmatrix}
0&0&60&0&0\\
0&50&0&0&80\\
0&50&60&71.6667&98.3333
\end{bmatrix}
\ {\rm MW}.
\label{eq:optimal_dispatch}
\end{equation}
The dispatch satisfies the demand in every period, since the row sums are
60, 130, and 280 MW, respectively. The resulting operating cost is
\$12,678 for all three modes, corresponding to a zero optimality gap relative to brute force.
\begin{table}[t]
\centering
\caption{Comparison of Block~2 Solver Modes}
\label{tab:solver_results}
\begin{tabular}{lcc}
\hline
\textbf{Solver Mode} & \textbf{Bitstring} & \textbf{Cost (\$)} \\
\hline
Brute force
& 001000100101111 & 12678.45  \\
Monolithic QAOA
& 001000100101111 & 12678.45  \\
Distributed QAOA
& 001000100101111 & 12678.45  \\
\hline
\end{tabular}
\end{table}
These results demonstrate two aspects of the proposed framework. First,
replacing brute-force enumeration with monolithic QAOA or Distributed QAOA in Block~2 does not change the recovered commitment bitstring, dispatch, operating cost, or ADMM convergence for the tested case. Second, Distributed QAOA obtains the same solution by distributing the 15 logical commitment qubits across multiple QPUs, avoiding the requirement that a single QPU contain the complete binary problem. The result therefore demonstrates solution consistency and multi-QPU capacity accommodation.

\section{Conclusion}
\label{sec:conclusion}

This paper developed a DQAOA-enabled three-block ADMM framework for unit commitment by integrating the DQAOA package into the binary QUBO block while keeping the remaining ADMM updates unchanged. The DQAOA interface enables solving the QUBO using brute-force enumeration, monolithic QAOA, or distributed QAOA with identical UC data, ADMM parameters, and convergence criteria. For the five-unit, three-period instance, brute-force enumeration, monolithic QAOA, and distributed QAOA all converged to a primal residual tolerance and recovered the same commitment schedule, dispatch, and operating cost of \$12,678. Distributed QAOA achieved this result by allocating the logical commitment qubits across multiple capacity-constrained QPUs, demonstrating solution consistency and the ability to accommodate multi-QPU capacity constraints. Future work will consider larger UC formulations and noisy or hardware-based implementations.

	\bibliographystyle{IEEEtran}
	\bibliography{references}

\end{document}